\documentclass[]{interact}

\usepackage{epstopdf}
\usepackage{subfigure}
\usepackage{graphicx}
\usepackage[latin1]{inputenc}
\usepackage{multirow}

\usepackage[numbers,sort&compress]{natbib}
\bibpunct[, ]{[}{]}{,}{n}{,}{,}
\makeatletter
\def\NAT@def@citea{\def\@citea{\NAT@separator}}
\makeatother

\theoremstyle{plain}

\theoremstyle{definition}

\theoremstyle{remark}

\begin{document}

\title{Statistical and Computational Tradeoff in Genetic Algorithm-Based Estimation}

\author{
	\name{Manuel Rizzo\textsuperscript{a}*\thanks{*Corresponding Author. Email: manuel.rizzo@uniroma1.it} and Francesco Battaglia\textsuperscript{a}}
	\affil{\textsuperscript{a}Department of Statistical Sciences, Sapienza University of Rome, Piazzale Aldo Moro 5, I-00185 Rome, Italy}
}

\maketitle

\begin{abstract}
	When a Genetic Algorithm (GA), or a stochastic algorithm in general, is employed in a statistical problem, the obtained result is affected by both variability due to sampling, that refers to the fact that only a sample is observed, and variability due to the stochastic elements of the algorithm. This topic can be easily set in a framework of statistical and computational tradeoff question, crucial in recent problems, for which statisticians must carefully set statistical and computational part of the analysis, taking account of some resource or time constraints. In the present work we analyze estimation problems tackled by GAs, for which variability of estimates can be decomposed in the two sources of variability, considering some constraints in the form of cost functions, related to both data acquisition and runtime of the algorithm. Simulation studies will be presented to discuss the statistical and computational tradeoff question.
\end{abstract}


\section{Introduction}

\label{intro}
In recent years the huge growth in size of datasets has introduced many novel problems in the statistical field. In fact the need to carry out successful statistical analysis must now be accompanied by a careful setting of the computational part, that may include the choice of computational methodology and must consider some resource or time constraints, that are crucial in real problems. Questions like these are known in literature as \textit{statistical and computational tradeoff} (or time-data tradeoff) problems, that aim at balancing and optimizing statistical efficiency and computational complexity. This is a very general topic, so many different methodologies have been proposed in literature to deal with many different applications. Chandrasekaran and Jordan \cite{chandra} considered a class of parameters estimation problems for which they studied a theoretical relationship in the form of a convex relaxation between number of statistical observations, runtime of the selected algorithm and statistical risk. An algebraic hierarchy of these convex relaxations is built to successfully achieve the time-data tradeoff for different algorithms. Dillon and Lebanon \cite{dillon} studied consistency of intractable Stochastic Composite Likelihood estimators, whose formula depends also on parameters related to computational elements. Therefore they aimed at balancing statistical accuracy and computational complexity. Shender and Lafferty \cite{shender} studied the tradeoff in Ridge Regression models introducing sparsity in the sample covariance matrix. Wang et al. \cite{wang}, in a Sparse Principal Component Analysis framework, addressed the question of whether is possible to find an estimator that is computable in polynomial time, and then they analyzed its minimax optimal rate of convergence. Several other applications can be found in \cite{jordan,berthet,bruer,chen,agarwal}.

In the present paper we address the statistical and computational tradeoff discussion in complex models building problems to be optimized by Genetic Algorithms (GAs), in a pure statistical approach. They have been widely employed in statistical applications \cite{baragona,kapanoglu,rizzo,satman,gurunlu}, but the question could be discussed also considering other evolutionary methods. The key point, in fact, is that algorithms including stochastic element introduce an additional source of variability in the estimation process along with variability induced by sampling: statistical efficiency of the estimates will be evaluated by considering the effect of both of these components. The tradeoff will finally be discussed by introducing some cost functions in the analysis related to both data acquisition and runtime of the algorithm. The applications considered are a Linear Regression model to be estimated by \textit{Least Absolute Deviation}, an \textit{Autoregressive} model simultaneous identification and parameter estimation, and a \textit{g-and-k distribution} maximum likelihood estimation, a kind of so-called \textit{intractable likelihood} problem.

The paper is organized as follows: Section \ref{sec:GAs} describes standard GAs and their application in parameters estimation problems; in Section \ref{sec:prob_descr} the tradeoff question is discussed, along with a literature review on GA variability quantification; Section \ref{sec:applic} illustrates the selected applications, for which the tradeoff is analyzed; last section includes final comments and future developments.

\section{Genetic Algorithms for Models Building}
\label{sec:GAs}

\subsection{Overview of the Algorithm}
Genetic Algorithms (GAs) are among the most important methodologies in the Evolutionary Computation field, because of their simplicity and versatility of applications. They were introduced by Holland \cite{holland} as a method to explain the adaptive processes of natural systems, using metaphors from biology and genetics, but soon they found application in complex optimization problems. The complexity may be due to the objective function, that may be non-differentiable for example, or to the search space, possibly very large or irregular.

In this framework the goal is to find the global optimum of a function, called \textit{fitness}, that measures the goodness of solutions. In the metaphor every
solution is represented by an individual, coded in a string called \textit{chromosome}, whose elements represent the genetic heritage of the individual (\textit{genes}). In the standard binary coding case genes can only take values $0$ or $1$ (\textit{bits}). In every iteration (or \textit{generation}, in the terminology of GAs) the algorithm considers a population of individuals of fixed size $N$ that evolves by use of genetic operators of \textit{selection}, \textit{crossover} and \textit{mutation}. The \textit{selection} randomly chooses solutions for the subsequent steps, usually proportionally to their fitness value; by \textit{crossover} two solutions are allowed to combine together, with a fixed rate $pC$, exchanging part of their genes, creating two new individuals; lastly, the \textit{mutation} step allows every bit to flip its value from 0 to 1, or vice versa, with a fixed probability $pM$, providing a further exploration of different areas of the search space. The resulting population replaces the previous, and the flow of generations stops if a certain condition is met, for example a fixed number of generations. It is also possible, adopting the \textit{elitist} strategy, to maintain the best individual found up to the current generation, in spite of the effect of genetic operators. In this case, the user interested in optimization may consider just the flow of these solutions.

Elitism is crucial as far as convergence is contemplated. In fact most of convergence results have been obtained for elitist GAs, generally by use of Markov Chain theory. A fundamental theorem by Rudolph \cite{rudolph}, easily adaptable to a wide class of Evolutionary Algorithms (EAs), considers an elitist GA with $pM>0$ and models $X_g$, namely the best solution found up to generation $g$, by a Markov Chain. It states that under the above assumptions the sequence $D_g=f^* - f(X_g)$, where $f^*$ is the global optimum and $f(X_g)$ the fitness of the best solution found up to generation $g$, is a non-negative supermartingale which converges almost surely to zero. Generalizations have been proposed to extend Rudolph's approach to time varying mutation and/or crossover rates by modeling GA as a non homogeneous Markov Chain \cite{rojascruz,pereira_deandr,pereira_campos}. Reference \cite{pereira_campos} includes also a review of other ways of studying GAs convergence by Markov Chain modeling. In our paper we employ a simple GA, so we shall mainly refer to Rudolph's theorem of convergence, that allows also to easily generalize the framework to other EAs. It is worth noting that this theorem just states the convergence of a GA, but it gives no information about its rate.

\subsection{Statistical Parameters Estimation}
\label{subsec:GA_config}
There are many statistical problems where complexity is high, for example \textit{outliers detection}, \textit{cluster analysis} or \textit{design of experiment}. Here we consider parametric model building problems, where the function to be maximized, a likelihood for example, is hardly tractable, and standard methods may fail in finding good estimates. In this situation a sample $\underline{y}$ is generated from a distribution known up to a parameters vector $\underline{\theta}$. The inference on $\underline{\theta}$ is made by maximizing an objective function that depends on both $\underline{\theta}$ and $\underline{y}$. 

We shall now specify GA solutions coding and fitness function structure. Even if floating-point GAs have been employed in literature to deal with real parameters optimization we shall employ the simple binary coded GA described above. The standard rule to binary code a real parameter $\theta$ in the real interval $[a,b]$ is:

\begin{equation*}
\theta=a+\frac{b-a}{2^{H}-1}\sum_{j=1}^{H}2^{j-1}x_{j}\thinspace,
\end{equation*}

where $H$ is the number of genes considered and $x_{j}$ is the $j$-th bit. If the interest is focused on a vector $\underline{\theta}=(\theta_1,...,\theta_k)$ then a chromosome of length $M=k\cdot H$ includes the coding of each component. Length $H$ of each genes group is constant, but the coding interval $[a,b]$ can vary for each parameter. Since we are considering a kind of discretization of a continuous search space, we aim at building a fine grid in such a way that fitness function is adequately smooth on that grid, so that the related loss of information is negligible.

The fitness $f$ is proportional to the above-cited objective function, say $g(\underline{\theta};\underline{y})$. We shall consider a scaled exponential transformation of $g$:

\begin{equation}
f(\underline{\psi})=\exp\{g(\underline{\theta};\underline{y})/\tau\}\thinspace,\label{eq:fitness}
\end{equation}

where $\underline{\psi}$ is the chromosome and $\tau$ is a problem dependent constant. This kind of scaling procedure may allow to modify the shape of fitness function without changing solutions ranking.

As far as the choice of genetic operators is considered, we shall adopt: \textit{roulette wheel selection}, to select chromosomes proportionally to their fitness value; \textit{single point crossover}, so that a chromosome can exchange up to $k-1$ parameters in every recombination; standard \textit{bit-flip mutation} strategy. Lastly, \textit{elitism} is adopted to guarantee convergence of the procedure.

\section{Problem Description}
\label{sec:prob_descr}

\subsection{Variability Decomposition}
\label{var_decomp}
A statistical parameter estimate is naturally subject to sampling variability; in fact if we make inference using two different samples we may obtain two possibly different results. This issue had to be deepened in all statistical inference approaches; in the present work we consider a classical approach, where sampling variability is closely related to variability of selected estimators.

When GAs are employed in the estimation process a new form of variability is introduced in the analysis, due to the stochastic nature of the algorithm. It refers to elements like the starting population, selection mechanism, mutation and crossover rate or the random choice of the cutting point of crossover. As a result of this, if we run a GA several times using the same sample we may obtain different results.

The total variability of a GA estimate can be easily decomposed in these two forms of variability, as shown in \cite[p. 50]{baragona} for the univariate case.

We shall adopt the following notation: $\underline{y}$ is the sample of observations, $\theta$ the parameter of the generating statistical model, $\widehat{\theta}(\underline{y})$ the best theoretical value, that can not be computed in practice, and $\theta^*(\underline{y})$ the result of optimization obtained via GA, that is an approximation of $\widehat{\theta}(\underline{y})$ and depends from the observed sample as well. The error of a GA estimate, under the assumption of independence between data-generating model and random seeds of the GA, can be decomposed as:

\begin{equation}\label{eq:decomptheta}
\theta^*(\underline{y}) -\theta = [\widehat{\theta}(\underline{y}) - \theta] + [ \theta^*(\underline{y})-\widehat{\theta}(\underline{y})] 
\end{equation}

The first term in square brackets depends on consistency of the estimates, while the second depends on the convergence of GA. Both must be ensured to allow the GA estimate to converge to zero in probability. A similar issue has been analyzed in \cite{winker}, where a Threshold Accepting algorithm is employed for a GARCH model estimation problem.

As long as we focus on models indexed by a vector $\underline{\theta}=(\theta_1,...,\theta_k)$ then in practice we shall consider the corresponding multiparametric of (\ref{eq:decomptheta}). This means that we must define two random vectors $\underline{\widehat{\theta}}(\underline{y})$ and $\underline{\theta}^*(\underline{y})$, which are affected, respectively, by sampling variability and GA variability. While $\underline{\widehat{\theta}}(\underline{y})$ is defined as the best statistical estimator, the GA component, for which the sample $\underline{y}$ is held fixed, needs to be defined specifically for our own estimation problem. 

If an elitist strategy is employed then we define the random vector $\underline{\theta}^{*(g)}(\underline{y})$ as the best estimate obtained up to generation $g$, that corresponds to the best individual of generation $g$. The idea is to evaluate GA variability considering the behaviour of this random vector among GA runs, and the key is Rudolph's theorem. In fact it states that if, along with elitist strategy, mutation rate is greater than zero, then the sequence $\underline{\theta}^{*(g)}(\underline{y})\hspace{4pt}(g=0,1,...)$ will converge almost surely to the optimum, that is $\underline{\widehat{\theta}}(\underline{y})$ in our case. This means that when $g$ increases then each GA run is likely to approach the optimum, so variability between runs tends to decrease as well. So, in our framework, evaluating the variability of the GA coincides with studying the convergence rate of the algorithm.

Having defined both random vectors $\underline{\widehat{\theta}}(\underline{y})$ and $\underline{\theta}^*(\underline{y})$ we shall define their variance-covariance matrices, respectively $\Sigma_{S}$ and $\Sigma_{GA}$, to relate to (\ref{eq:decomptheta}).
The generic $(i,j)$ elements of these matrices are:

\begin{equation*}
\sigma^{S}_{ij}= \mathbb{E}_{S}[ (\widehat{\theta}_i-\theta_i) (\widehat{\theta}_j-\theta_j)],\hspace{6pt}i,j=1,...k,
\end{equation*}

\begin{equation*}
\sigma^{*}_{ij}=\mathbb{E}_{GA}[ (\theta_{i}^*-\widehat{\theta}_i) (\theta_{j}^*-\widehat{\theta}_j)],\hspace{6pt}i,j=1,...k.
\end{equation*}

So $\sigma^{S}_{ij}$ and $\sigma^{*}_{ij}$ measure the dependence between $\theta_i$ and $\theta_j$ induced, respectively, by sampling and GA. As long as we need to get a scalar summary of these matrices, a possible choice is to consider the traces only, a strategy generally adopted in literature. This makes a good sense in an optimization framework, because the optimum is reached when variances $\sigma^{S}_{ii}$ and $\sigma^{*}_{ii}$ ($i=1,...,k$) go to zero, with no practical interest on covariances. So, if $\Sigma_{TOT}$ is the total variance-covariance matrix, then, using the linearity of trace, and under the same independence assumption of (\ref{eq:decomptheta}), we can write:
\begin{equation}\label{eq:decompSigma}
tr(\Sigma_{TOT})=tr(\Sigma_{S})+tr(\Sigma_{GA}).
\end{equation}

\subsection{Tradeoff Problem}
\label{subsec:tradeoff}

Now we shall set the variability analysis of subsection \ref{var_decomp} in the framework of statistical and computational tradeoff. Assuming that both statistical estimator and GA's configurations are fixed, then we must figure out how to optimally balance statistical accuracy and GA efficiency.

As long as we consider estimators having property of consistency, then statistical accuracy can be naturally represented by the size $n$ of sample $\underline{y}=(y_1,...,y_n)$, because if $n$ increases then also the precision of estimators increases (and, in contrast, variability decreases). This happens under some regularity conditions (see \cite[p.470]{casella}, in the case of Maximum Likelihood Estimators).

Concerning GA efficiency, we refer to Rudolph's theorem of convergence. Informally, a GA converges when $g$ tends to infinity, but it is worth noting that in every GA generation each of the $N$ chromosomes in the population is evaluated on the basis of fitness function. So, instead of considering the number of generations, we represent GA efficiency component by the number of fitness function evaluations $V$, also because it is usually the most computationally expensive step.

Since we want to figure out if both $tr(\Sigma_{S})$ and $tr(\Sigma_{GA})$ go to zero, we shall study their behaviour when $n \to \infty $ and $V \to \infty $. Let us consider two functions $f(n)$ and $h(V)$ for which, respectively, $f(n) \to \infty$ when $n \to \infty$ and $h(V) \to \infty$ when $V \to \infty$. If we consider a consistent statistical estimator and the assumptions of Rudolph's theorem are fulfilled, then we can write $tr(\Sigma_{S})=\mathcal{O}([f(n)]^{-1})$ and $tr(\Sigma_{GA})=\mathcal{O}([h(V)]^{-1})$. In that case:

\begin{equation} \label{eq:tr_sigmatot}
tr(\Sigma_{TOT})=tr(W_{S})\frac{1}{f(n)} + tr(W_{GA})\frac{1}{h(V)},
\end{equation}

where matrices $W_{S}$ and $W_{GA}$ are constant and composed by elements that depend, respectively, from the statistical model and from the GA. It is possible that sample size $n$ may have an effect also on $W_{GA}$, because fitness function will change as a consequence. For this reason we shall include $n$ in our fitness scaling procedure as constant $\tau$ in (\ref{eq:fitness}). In this way we can strongly restrict the effect of $n$ on behaviour of the algorithm and describe the total variability of a GA estimate by considering decomposition (\ref{eq:tr_sigmatot}).

To specify the total effort of a GA estimate we shall introduce some cost functions: $S(n)$ represents the cost for obtaining a sample of $n$ observations, while $T(n)$ indicates the cost of one fitness function evaluation, which depends on the number of observations as well, because a solution is evaluated by analyzing the full sample. So the total cost $C$ for obtaining an estimate $\theta^*(\underline{y})$ using $n$ statistical observations and $V$ fitness function evaluations is given by: $C=S(n)+VT(n)$.

Now we can write our tradeoff question as an optimization problem:

\begin{equation*}
\begin{Bmatrix}
\underset{n,V}{min}\hspace{3pt}tr(\Sigma_{TOT})=tr(W_{S})\frac{1}{f(n)} + tr(W_{GA})\frac{1}{h(V)}\\ 
s.t.\\ 
C=S(n)+VT(n)
\end{Bmatrix}
\end{equation*}

A particular case that simplifies the latter is the assumption of linearity in $n$ for cost functions $T$ and $S$. This is reasonable because statistical observations are usually collected in sequence and if GA's fitness function includes a summation over the considered sample. In such a case $T(n)=nT$, $S(n)=nS$ and we can incorporate the effort constraint into the objective function to obtain:

\begin{equation*}
\underset{n}{min}\hspace{3pt}tr(\Sigma_{TOT})=tr(W_{S})\frac{1}{f(n)} + tr(W_{GA})\frac{1}{h([C-nS]/nT)}.
\end{equation*}

A solution can be found numerically once consistency and convergence rates  $f(\cdot)$ and $h(\cdot)$ have been established. A particular case that allows to obtain a simple closed form expression for optimal $n$ and $V$ is given when $f(n)=n$ and $h(V)=V$. In this case, computing the derivative of the objective function with respect to $n$, we obtain solutions:

\begin{equation} \label{eq:nOPT}
\underline{\tilde{n}}=\frac{-S\hspace{1pt}C\hspace{1pt}tr(W_{S})\pm C\sqrt{C\hspace{1pt}T\hspace{1pt}tr(W_{S})\hspace{1pt}tr(W_{GA})}}{C\hspace{1pt}T\hspace{1pt}tr(W_{GA})-S^2\hspace{1pt}tr(W_{S})}.
\end{equation}

Since $n$ is a sample size, then we are interested only in the positive solution $\tilde{n}$ of (\ref{eq:nOPT}). $\tilde{V}$ is obtained by constraint:
\begin{equation} \label{eq:VOPT}
\tilde{V}=\frac{C-\tilde{n}\hspace{1pt}S}{\tilde{n}\hspace{1pt}T}.
\end{equation}

\subsection{Consistency and Convergence Rates}
\label{subsec:cons_conv}
Functions $f(n)$ and $h(V)$ introduced in the previous subsection specify, respectively, consistency rate of statistical part and convergence rate of algorithmic part of equation (\ref{eq:decompSigma}). The assumption of linearity is a particular case that simplifies the tradeoff analysis.

Linearity of $f(n)$ is satisfied if we consider estimators having property of asymptotic efficiency: in that case, under some regularity conditions, $f(n)=n$ (see \cite[p.472]{casella}, in the case of Maximum Likelihood Estimators).

On the other side, the behaviour of $h(V)$, as said, is related to the convergence rate of GAs in our case. This is an essential issue for any optimization algorithm, and in the field of Evolutionary Algorithms (EAs) has been analyzed in different ways. A part of literature focuses on comparison of EAs with different configurations, to identify the algorithm that reaches the optimum more quickly \cite{eiben,derrac}; other researchers have developed more rigorous approaches, focusing among other things on convergence rate of single chromosome bits, relatively to classic problems like \textit{OneMax} \cite{oliveto,auger}; a different proposal \cite{prugel,shapiro,reeves}, inspired by \textit{statistical mechanics}, aims to model the GA as a complex system, and to summarize its probability distribution through generations by considering cumulants. In such a way GA convergence can be evaluated by considering the limiting cumulants.

Recently Clerc \cite[p.69]{clerc} has proposed a theoretical framework for analyzing optimization performances. For a general stochastic algorithm (deterministic algorithms are considered particular cases of this class) he introduced a bivariate probability density $p(\psi,r)$, called \textit{Eff-Res}, that is function of both optimization \textit{result} $r$ and computational \textit{effort} $\psi$, spent for obtaining $r$. By analyzing this function it is possible to deepen different useful questions: for a given result $r$, the probability of obtaining $r$ with a generic effort $\psi$; for a given effort $\psi$, the probability of obtaining a generic result $r$. Our interest is focused on the latter question, because if we fix a computational effort, that is related to the number of fitness evaluations in our case, then we are interested in how the result $r$ varies. The theoretical variance of results for fixed effort can be written as:

\begin{equation} \label{eq:var.GA.Clerc}
\sigma^2(\psi)=\mu(\psi) \int_{\tilde{R}}(r-\bar{r}(\psi))^2 \hspace{2pt} p(\psi,r) dr,
\end{equation}

where $\tilde{R}$ is the set of possible results, $\bar{r}(\psi)$ the theoretical mean result for fixed effort and $\mu(\psi)$ the normalization coefficient of $p(\psi,r)$.
Expression (\ref{eq:var.GA.Clerc}) can be evaluated empirically. If we have observed $J$ results $r(1),r(2),...,r(J)$, obtained with effort $\psi$, then the estimated variance is given by:

\begin{equation} \label{eq:var.stim.GA.Clerc}
\hat{\sigma}^2(\psi) = \frac{1}{J-1} \sum_{j=1}^{J}[r(j)-\bar{r}_J(\psi)]^2,
\end{equation}

where $\bar{r}_J(\psi)$ is the empirical mean of results.

In the present work we shall employ a very similar approach to evaluate GA variability. Since we are interested in convergence of $\theta_{i}^*$ to the optimum $\widehat{\theta}_i$ $(i=1,...,k)$, then in both (\ref{eq:var.GA.Clerc}) and (\ref{eq:var.stim.GA.Clerc}) we plug $\widehat{\theta}_i$ in place of theoretical and empirical mean, and $\theta_{i}^*$ in place of results. In that case (\ref{eq:var.GA.Clerc}) corresponds to variance $\sigma^{*}_{ii}=\mathbb{E}_{GA}[ (\theta_{i}^*-\widehat{\theta}_i)^2]$ of matrix $\Sigma_{GA}$. If we run a GA $J$ times, obtaining $\theta_{1,i}^*,\hspace{1pt}\theta_{2,i}^*,\hspace{1pt}...,\theta_{J,i}^*\hspace{2pt}(i=1,...,k)$, thus we get the estimates by:

\begin{equation}
\label{eq:sigmaGA}
\hat{\sigma}_{ii}^{*} = \frac{1}{J} \sum_{j=1}^{J}[\theta_{j,i}^*-\widehat{\theta}_i]^2,\hspace{4pt}i=1,...,k.
\end{equation}

As far as we need quantifications of convergence rate, then we shall analyze the behaviour of these variances through generations. Thus we consider the sequence of variances for the $k$ dimensional parameter $\underline{\theta}$, given a fixed maximum number of generations G:

\begin{equation}
\label{eq:sigmaGA_g}
\underline{\hat{\sigma}}^{*(g)}=(\hat{\sigma}^{*(g)}_{11},\hat{\sigma}^{*(g)}_{22},...,\hat{\sigma}^{*(g)}_{kk}),\hspace{10pt}g=1,...,G.
\end{equation}

We shall now conduct the following regression analysis for each parameter indexed by $i$:

\begin{equation}
\label{eq:regr_sigmaGA}
\hat{\sigma}_{ii}^{*(g)}=w_{GA,i} \hspace{1pt} \frac{1}{[V^{(g)}]^a}+\epsilon_g, \hspace{10pt}g=1,...,G,
\end{equation}

where $[V^{(g)}]^a$ is the $a$-th power of the number of fitness evaluations up to generation $g$ and $w_{GA,i}$ is the regression parameter. The goal is to find out if exists an $a$ for which $[V^{(g)}]^a$ can be taken as a satisfactory GA convergence rate $h(V)$. In that case $w_{GA,i}$ will become part of matrix $W_{GA}$ in (\ref{eq:tr_sigmatot}); for this reason $a$ shall be the same for all parameters in the considered model.

\section{Applications}
\label{sec:applic}
The applications selected in this paper are a Least Absolute Deviation Regression estimation (code $LAD$), an Autoregressive model building (code $AR$) and a $g$-and-$k$ distribution maximum likelihood estimation (code $gk$). In order to discuss the tradeoff question for each of these experiment we shall now give details on methods employed for obtaining variability and convergence rate estimates, motivations on choices of estimators and GAs implementation. Simulations and computations were implemented by use of software R \cite{rcore} for all applications, and also R package \textit{gk} \cite{gk} for the last application.

Concerning GAs configuration we adopted coding, fitness and genetic operators described in subsection \ref{subsec:GA_config}, and coding interval boundaries specific for each parameter. Crossover and mutation rates were fixed at, respectively, $0.7$ and $0.1$, maximum number of generations $G$ at $1400$ and population size $N$ will be equal to $50$. If not otherwise specified the initial population was generated uniformly at random. These configurations have been chosen on the basis of empirical studies to guarantee stability and convergence of the procedure.

As mentioned in subsection \ref{subsec:cons_conv}, if an estimator is asymptotically efficient then $f(n)=n$ in formula (\ref{eq:tr_sigmatot}). We considered estimators which have this property. We then estimated variability of estimators by simulating $10000$ samples and computing mean squared deviations of estimates obtained by software optimization routines from the true parameters, to get a quantification of $W_{S}$ in (\ref{eq:tr_sigmatot}).

On the other side, GA variability have been estimated by considering $10$ equally-sized datasets. For each sample we computed variance estimates using $J=500$ GA runs as shown in formulas (\ref{eq:sigmaGA}) and (\ref{eq:sigmaGA_g}); then we considered point by point average of these estimates with respect to $g$, obtaining final estimates to conduct regression analysis (\ref{eq:regr_sigmaGA}).

\begin{figure}
	\centering
	\includegraphics[width=13cm]{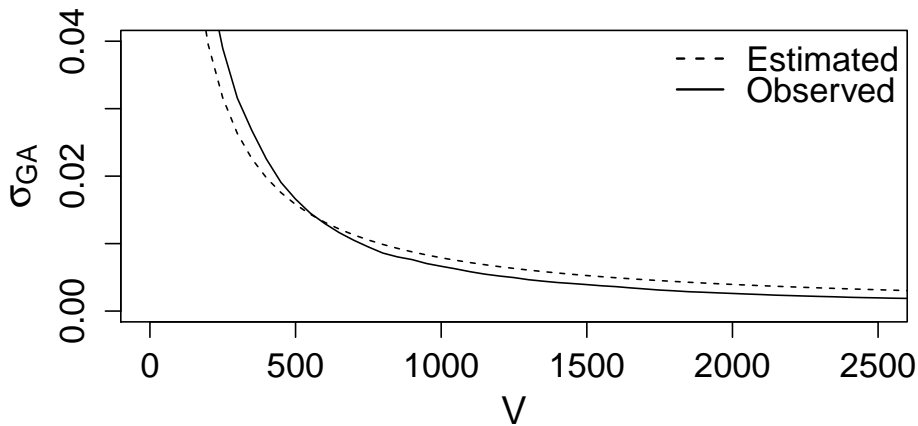}
	\caption{Observed (thick line) and estimated (dashed line) GA variability for parameter $\beta_1$ of $LAD$ experiment ($w_{GA}=7.9$, $R^2=0.97$)}
	\label{fig:Figure1}
\end{figure}

This regression analysis has been conducted for the three applications with $a=\frac{1}{3},\frac{1}{2}, 1,2$, and goodness of fit results ($R^2$ coefficient) are summarized in Table \ref{tab:rsquare}. Concerning experiments \textit{LAD} and \textit{gk} the best fits are observed for $a=1$, while $a=1/2$ rate is dominant for experiment \textit{AR}. We adopted these two convergence rates in the tradeoff analysis of next section. As an example Figure \ref{fig:Figure1} shows the fit for parameter $\beta_2$ of $LAD$ experiment.

\begin{table}
	\caption{$R^2$ coefficient values related to four different regression analysis conducted on each parameters of each experiment, in order to estimate convergence rate of $\Sigma_{GA}$}
	\label{tab:rsquare}
	\centering{}
	\vspace{5pt}
	\begin{tabular}{cccccc}
		\hline\noalign{\smallskip}
		Exp & Param & $a=1/3$ & $a=1/2$ & $a=1$ & $a=2$  \\
		\noalign{\smallskip}\hline\noalign{\smallskip}
		\multirow{3}{25pt}{$LAD$}
		& $\beta_0$ & 0.1883 & 0.4781 & 0.9775 & 0.7247 \\
		& $\beta_1$ & 0.1943 & 0.4835 & 0.9792 & 0.7298 \\
		& $\beta_2$ & 0.1910 & 0.4790 & 0.9763 & 0.7250 \\
		\noalign{\smallskip}\hline\noalign{\smallskip}
		\multirow{4}{20pt}{$gk$}
		& $A$ & 0.3538 & 0.6635 & 0.9525 & 0.6370 \\
		& $B$ & 0.2060 & 0.4949 & 0.9179 & 0.5984 \\
		& $g$ & 0.2722 & 0.5883 & 0.7585 & 0.3511 \\
		& $k$ & 0.1268 & 0.3563 & 0.9548 & 0.9071\\
		\hline\noalign{\smallskip}
		\multirow{8}{20pt}{$AR$}
		& $\phi_1$ & 0.7806 & 0.9338 & 0.8864 & 0.4655 \\
		& $\phi_2$ & 0.9101 & 0.9896 & 0.7083 & 0.2622 \\
		& $\phi_3$ & 0.9164 & 0.9835 & 0.6645 & 0.2200 \\
		& $\phi_4$ & 0.8998 & 0.9767 & 0.6762 & 0.2228 \\
		& $\phi_5$ & 0.8869 & 0.9726 & 0.6878 & 0.2306 \\
		& $\phi_6$ & 0.8801 & 0.9698 & 0.6921 & 0.2325 \\
		& $\phi_7$ & 0.8569 & 0.9597 & 0.7104 & 0.2453\\
		& $\phi_8$ & 0.8576 & 0.9635 & 0.7311 & 0.2641\\
		\noalign{\smallskip}\hline\noalign{\smallskip}
	\end{tabular}
\end{table}

\begin{table}
	\caption{Sampling and GA variability components estimates}
	\label{tab:matr_tr}
	\centering{}
	\vspace{5pt}
	\begin{tabular}{ccc}
		\hline\noalign{\smallskip}
		Exp & $tr(W_{S})$ & $tr(W_{GA})$  \\
		\noalign{\smallskip}\hline\noalign{\smallskip}		
		$LAD$ & 5.38 & 23.18\\
		$AR$ & 12.26 & 17.74\\
		$gk$ & 103.39 & 3897.25\\
		
		\noalign{\smallskip}\hline\noalign{\smallskip}
	\end{tabular}
\end{table}

Results of estimates of $tr(W_{S})$ and $tr(W_{GA})$ are summarized in Table \ref{tab:matr_tr}. For both estimates we used simulated data of length $n=200$ for all the experiments.

The tradeoff will be discussed for the three applications by evaluating optimal $\tilde{n}$ on a common grid of values for linear cost functions $S$ and $T$, assuming a fixed total effort $C=10^5$. Comments on optimal $V$ can be derived by complement. We shall make some remarks also for the case where we estimate computational cost $T$ with time (in seconds) needed in our computer to evaluate fitness in the three experiment, using $gk$ as corner point. In this way we can make more realistic comparative comments.

\subsection{\textit{Least Absolute Deviation} Estimation}
\label{subsec:LAD}
Least Absolute Deviations (LAD) regression is an alternative to Ordinary Least Squares regression, that has been proven to be more robust to outliers \cite[p.52]{bloomfield}. In this framework the estimator, that is asymptotically efficient \cite[p.44]{bloomfield}, is the function that minimizes the sum of absolute values of errors. This function is neither differentiable, nor convex, so numerical methods must be employed to find an optimal solution. Zhou and Wang \cite{zhou_wang} have already employed a real valued GA to estimate the parameters of a LAD regression with censored data. In this paper we consider a standard linear regression model:

\begin{equation*}
y_i= \beta_0 + \beta_1 x_{i,1} + \beta_2 x_{i,2} + \epsilon_i, \hspace{15pt}i=1,...,n,
\end{equation*}

where $(\underline{y},\underline{x})$ is the observed dataset and $\epsilon_i \sim t_5$.

Since our goal is always maximization, then the fitness function shall be:

\begin{equation*}
f(\underline{\psi})=\exp\{- \sum_{i=1}^{n} \left | y_i - \beta_0 - \beta_1 x_{i,1} - \beta_2 x_{i,2} \right | \hspace{2pt} / \hspace{2pt} n\}.
\end{equation*}

True parameters vector will be $\underline{\beta}=(0.5,0.5,-0.5)$, each chromosome length shall be $M=24$ and coding interval boundaries will be $[-2,2]$ for all parameters.

\begin{figure}[ht!]
	\centering
	\includegraphics[width=10cm]{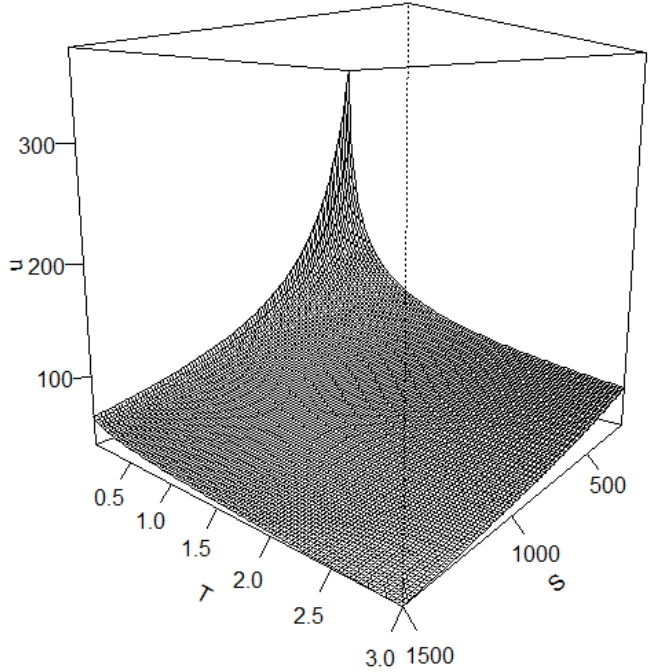}
	\caption{Behaviour of optimal $n$ for experiment \textit{LAD}}
	\label{fig:Figure2}
\end{figure}

Figure \ref{fig:Figure2} shows the behaviour of optimal $n$ (on z axis) with respect to a grid of values for cost functions $S$ and $T$.
It obviously increases to large values as costs $S$ and $T$ decrease, and rapidly decreases as they increase.

\subsection{\textit{Autoregressive} Models Building}
\label{subsec:AR}
GAs have been widely applied in the field of time series analysis. In fact optimization problems related to parameters estimation and model identification may have some difficulties due to the intractability of objective functions or to the size of search spaces. The latter question is common in model identification problems, and it has been analyzed also for standard ARMA models \cite{gaetan,minerva}. Here we address the problem of how to simultaneously identify and estimate Autoregressive (AR) models, given a fixed maximum order.

The general equation of an AR model of order $p$ is:
\begin{equation} \label{eq:AR_model}
Y_t=\phi_1 Y_{t-1} + ... + \phi_{p} Y_{t-p} + \epsilon_t,
\end{equation}

where $Y_t$ is a zero mean random process, $\epsilon_t$ a Gaussian white noise and $\underline{\phi}=(\phi_1,...,\phi_p)$ the parameters vector.

Model (\ref{eq:AR_model}) is usually identified by using penalized likelihood criteria like AIC or BIC, to be minimized. In this work we shall consider BIC, because of its property of consistency \cite{hannan}. As long as we need to simultaneously identify and estimate the model, we shall not include maximized likelihood in the criterion, but just the likelihood evaluated for a generic vector $\underline{\phi}$. Thus we get:

\begin{equation} \label{eq:BIC}
BIC(\underline{\phi};\underline{y})=n \hspace{2pt} log \hspace{1pt} \hat{\sigma}^2(p) + k \hspace{2pt} log \hspace{1pt} n,
\end{equation}

where $\underline{y}$ is the observed time series, $\hat{\sigma}^2(p)= \sum_{i=1}^{n} (y_t - \phi_1 y_{t-1} - ... - \phi_{p} y_{t-p})^2 / n$ and $k \leq p$ is the number of free parameters in the model. Sampling variability will be estimated considering asymptotic efficiency of maximum likelihood estimator for AR models \cite[p.386]{brockwell}.

True model is an $AR(1)$ with $\phi_1=0.8$, and we shall consider a maximum order $p=8$. Chromosome length shall be $M=64$, and coding will necessarily include the case $\phi_i=0$ ($i=1,...,8$), that has a direct impact on the penalization term of (\ref{eq:BIC}). To facilitate the identification of subset models we shall force the starting population to include a chromosome that corresponds to a white noise (all parameters are zero), and also $8$ chromosomes for which one of the parameters is zero, so that all $\phi_i=0$ ($i=1,...,8$) are represented; the remaining chromosomes will be generated uniformly at random, coherently with other applications. This may be a reasonable strategy in a situation of total lack of knowledge.

Fitness function shall be:

\begin{equation*}
f(\underline{\psi})=\exp\{- BIC(\underline{\phi};\underline{y}) \hspace{2pt} / \hspace{2pt} n\},
\end{equation*}

while coding interval will be $[-2,2]$ for each $\phi_i$.

\begin{figure}[ht!]
	\centering
	\includegraphics[width=10cm]{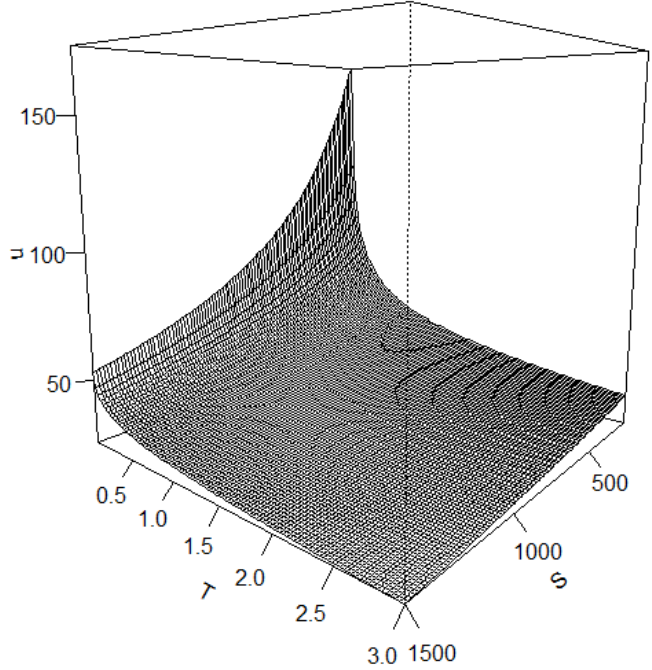}
	\caption{Behaviour of optimal $n$ for experiment \textit{AR}}
	\label{fig:Figure3}
\end{figure}

Figure \ref{fig:Figure3} shows the analogous plot to Figure \ref{fig:Figure2}. This experiment has slower GA convergence rate with respect to \textit{LAD} and \textit{gk}, possibly because of the effect of model identification in the fitness 
(e.g. estimating a $\phi_i$ value slightly different from zero implies may implies a slight decrease of the residual sum of squares but a term $k$ one unit larger in the penalization part of BIC). For this reason values of optimal $n$ are generally lower than \textit{LAD}.

\subsection{\textit{g-and-k} Distribution Estimation}
\label{subsec:gk}
The \textit{g-and-k} distribution was introduced by Haynes et al. \cite{haynes_gatton}, and is a family of distributions specified by a quantile function. It is a very flexible tool which has been applied in statistical control charts techniques \cite{haynes_meng_rippon} and non-life insurance modelling \cite{peters}. For a univariate random sample $\underline{x}=(x_1,...,x_n)$ the quantile function is:

\begin{equation*}
Q_X(u_i|A,B,g,k)=A + B \hspace{1pt} z_{u_i}  \left ( 1 + c\frac{1-e^{-g z_{u_i}}}{1+e^{-g z_{u_i}}} \right ) (1+z^2_{u_i})^k, \hspace{15pt}i=1,...,n,
\end{equation*}

where $u_i=F_X(x_i|A,B,g,k)$ is the depth corresponding to data value $x_i$, $z_{u_i}$ the $u_i$-th quantile of standard normal distribution, $A$ and $B>0$ are location and scale parameters, $g$ measures skewness in the distribution, $k>-0.5$ is a measure of kurtosis and $c$ is a constant introduced to make the distribution proper. By combining values of the four parameters several essential distributions like Normal, Student's t or Chi square can be derived.

Maximum Likelihood estimation of this distribution falls in the case of so called \textit{intractable likelihood} problems. The expression of likelihood is given by:

\begin{equation}\label{eq:gk_likel}
L(\underline{\theta} \hspace{1pt} | \hspace{1pt} \underline{x})=\left (  \prod_{i=1}^{n} Q'_X( Q^{-1}_X(x_i \hspace{1pt} | \hspace{1pt} \underline{\theta}) \hspace{1pt} | \hspace{1pt} \underline{\theta})  \right )  ^{-1},
\end{equation}

where $\underline{x}$ is the observed sample, $\underline{\theta}=(A,B,g,k)$ and $Q'_X( u \hspace{1pt} | \hspace{1pt} \underline{\theta})= \partial \hspace{1pt} Q_X / \partial u$.

The main complication in computing (\ref{eq:gk_likel}) is that there is no closed form for the expression $Q^{-1}_X(x_i \hspace{1pt} | \hspace{1pt} \underline{\theta})$, that must be obtained numerically, for example with Brent's method, commonly used in many softwares.

\begin{figure}[ht!]
	\centering
	\includegraphics[width=10cm]{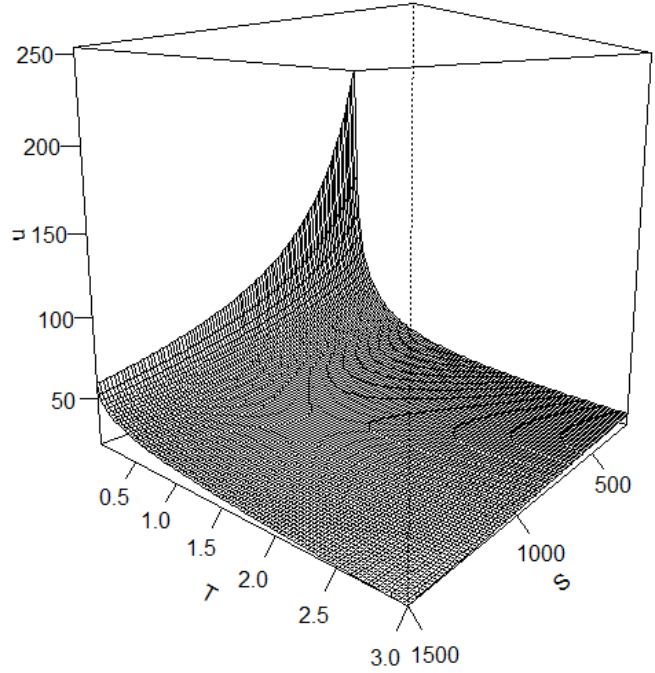}
	\caption{Behaviour of optimal $n$ for experiment \textit{gk}}
	\label{fig:Figure4}
\end{figure}

A lot of research on $g$-and-$k$ distributions estimation has been made in a Bayesian framework, using Markov Chain Monte Carlo \cite{haynes_meng} or indirect methods like Approximate Bayesian Computation \cite{allingham,grazian}.

In this paper we shall follow the pure likelihood approach proposed by Rayner and MacGillivray \cite{rayner}. In this situation a numerical procedure has to be selected to maximize (\ref{eq:gk_likel}). They proposed a Nelder-Mead simplex algorithm, reporting some weaknesses and highlighting the need to use several starting point for the optimization. In the final discussion they also observed that metaheuristic methods like GAs could be more successful in this optimization problem.

In our GA approach we shall consider the fitness:

\begin{equation*}
f(\underline{\psi})=\exp\{ \hspace{2pt} log \hspace{1pt} L(\underline{\theta} \hspace{1pt} | \hspace{1pt} \underline{x}) /n \hspace{2pt} \}.
\end{equation*}

We will simulate data using the typical parameters generator vector $\underline{\theta}=(A,B,g,k)=(3,1,2,0.5)$, with $c=0.8$, that leads to an 'interesting far-from-normal distribution' \cite[p. 192]{allingham}.

Each chromosome will have length $M=28$, and coding interval boundaries shall be $A \in [-10,10]$, $B \in [0,10]$, $g \in [-10,10]$ and $k \in [-0.5,10]$. If a decoded chromosome provides unacceptable values $B=0$ or $k=-0.5$ it is rejected and regenerated.

Concerning sampling variability, Rayner and MacGillivray \cite{rayner} investigated the approximation of maximum likelihood estimator variability by Cramer-Rao variance bound, which is of order $\mathcal{O}(n^{-1})$. In estimating sampling variability we shall allow for this asymptotic approximation of $\Sigma_{S}$.

Perspective plot for this experiment (Figure \ref{fig:Figure4}) shows a similar behaviour of optimal $n$ to $AR$, because even if in this case there is a linear GA convergence rate, the experiment is more complex ($tr(W_{GA})/tr(W_{S})$ ratio is much larger).

\begin{figure}[ht!]
	\centering
	\includegraphics[width=12cm]{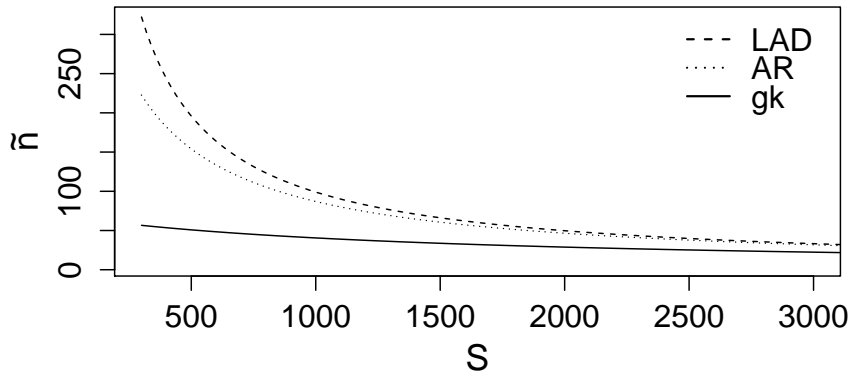}
	\caption{Optimal sample size with fixed estimated computational cost}
	\label{fig:Figure5}
\end{figure}

Lastly we shall make some comments on the behaviour of $\tilde{n}$ when sampling cost $S$ varies and fitness evaluation cost $T$ is estimated in each experiment by elapsed execution time (in seconds) of our computer for single fitness evaluation, taking $gk$ as corner point. Results are: $T_{LAD}/T_{gk}=0.007$ and $T_{AR}/T_{gk}=0.101$. Figure \ref{fig:Figure5} shows the behaviour of $\tilde{n}$ in this more realistic scenario, for which each computational cost ratio has been multiplied by a constant to highlight the behaviour of each experiment.
In this case the three curves are ranked with respect of computational cost and experiment complexity, that is related on both GA convergence rate and variability ratio $tr(W_{GA})/tr(W_{S})$ magnitude. $gk$ experiment shows lowest values of $\tilde{n}$, but when $S$ increases the three experiments tend to conform to common values, suggesting that a large sampling cost could have a larger influence in the tradeoff than model complexity.

\section{Discussion and Further Developments}
\label{sec:discussion}
In this paper we proposed a statistical and computational tradeoff analysis for complex estimation problems tackled by GAs based on a decomposition of variability of estimates in two terms depending, respectively, on sampling and stochastic features of the algorithm. Results of applications showed how the behaviour of optimal sample size changes with complexity of experiment. A comparative analysis of the three experiments in which computational cost is estimated also suggested that large sampling cost could influence optimal values more than complexity of the model, represented by statistical and computational variability. This is an interesting consideration, especially for real applications, where often large costs can decisively restrict the analysis.

The present study could be improved by considering other scalar quantifications of statistical and computational variability. For example one could consider the determinant of $\Sigma_{S}$ and $\Sigma_{GA}$ instead of trace. An other direction for further research is to generalize this framework to other statistical problems where a GA is involved. In fact, as already said, there are many complex optimization problems in the statistical field, and going deeply through the understanding of the tradeoff could facilitate integration of GAs in standard statistical methods. Lastly the discussion of statistical and computational tradeoff could be interesting also in estimation problems when nature inspired algorithms for continuous optimization are employed, like \textit{Differential Evolution} (DE) \cite{price} or \textit{Particle Swarm Optimization} (PSO) \cite{kennedy}, for which there is direct real coding. In fact the specific stochastic elements in these algorithms, for example the differential mutation mechanism in DE or the parameter regulating particle velocity in PSO, could result in different convergence rates for the algorithmic variability.

\section*{Acknowledgements}
This research has been financially supported by Sapienza University of Rome (grant \textit{Progetti per Avvio alla Ricerca} 2016).

\end{document}